\documentstyle[12pt,epsfig]{article}

\def\t{\tilde}
\def\D{\Delta}
\def\L{\Lambda}
\def\e{\epsilon}
\def\be{\begin{equation}}
\def\ee{\end{equation}}
\def\bea{\begin{eqnarray}}
\def\eea{\end{eqnarray}}

\newcommand{\AmS}{{\protect\the\textfont2
\renewcommand{\thesection}{\Roman{section}}
  A\kern-.1667em\lower.5ex\hbox{M}\kern-.125emS}}

\begin{document}

\vskip 2. truecm
\centerline{\bf Probability Distribution Function of the Diquark
Condensate}
\centerline{\bf in Two Colours QCD}

\vskip 2 truecm
\centerline { R. Aloisio$^{a,d}$, V.~Azcoiti$^b$, G. Di Carlo$^{c,d}$, 
A. Galante$^{a,d}$ and A.F. Grillo$^d$}
\vskip 1 truecm
\centerline {\it $^a$ Dipartimento di Fisica dell'Universit\`a 
di L'Aquila,  67100 L'Aquila, (Italy).}
\vskip 0.15 truecm
\centerline {\it $^b$ Departamento de F\'\i sica Te\'orica, Facultad 
de Ciencias, Universidad de Zaragoza,}
\centerline {\it 50009 Zaragoza (Spain).}
\vskip 0.15 truecm
\centerline {\it $^c$ Istituto Nazionale di Fisica Nucleare, 
Laboratori Nazionali di Frascati,}
\centerline {\it P.O.B. 13 -  00044 Frascati, (Italy). }
\vskip 0.15 truecm
\centerline {\it $^d$ Istituto Nazionale di Fisica Nucleare, 
Laboratori Nazionali del Gran Sasso,}
\centerline {\it  67010 Assergi (L'Aquila), (Italy). }
\vskip 3 truecm

\centerline {ABSTRACT}
\vskip 0.5truecm

\noindent
We consider  diquark condensation in finite density
lattice SU(2).
We first present an extension of Vafa-Witten result, 
on spontaneous breaking of 
vector-like global symmetries, that allows us to formulate a 
no-go theorem for diquark condensation in a region of the chemical 
potential-mass parameter space.
We then describe a new technique to calculate diquark condensation
at any number of flavours directly at zero external source 
without using any potentially dangerous extrapolation procedure.
We apply it to the strong coupling limit and
find compelling evidences for a second order phase
transition, where a diquark condensate appears, as well as
quantitative agreement between lattice results
and low-energy effective Lagrangian calculations.

\vfill\eject

\section{Introduction}

\vskip 0.3truecm

Recently the expected  scenario for the phase diagram of QCD
in the chemical potential-temperature plane has changed:
besides the hadronic and quark-gluon plasma phases,
the existence of a new state of matter has been proposed by several 
groups\cite{wil}.
This new phase is characteristic of the high density, low temperature 
regime: the asymptotic freedom of QCD and the known instability
of large Fermi spheres, in presence of (whatever weak) attractive forces,
results in a pairing of quarks with 
momenta near the Fermi surface   (in analogy with the 
Cooper pairing in solid state systems at low temperature).
A condensation of quark pairs should be the distinctive signal of the
new phase and has been indeed predicted using simplified 
phenomenological models of QCD.

Unfortunately the lattice approach, the most powerful tool to perform
first principles, non perturbative studies, is affected in the case of 
finite density QCD by the well known sign problem that has prevented
until now any step towards the understanding of this new phase.
This is not the case for the SU(2) theory where the fermions are in the
pseudo-real representation and finite density numerical simulations 
are feasible.

In this paper we present a detailed study of diquark condensation
for the unquenched two colours model. 
This work follows a previous one
where we considered chiral and diquark susceptibilities and found
strong evidence for a phase transition separating the ordinary
low density phase from a high density one where the chiral condensate
vanishes and the baryon number symmetry is broken\cite{noi1}.
In the present work we mainly focus on an approach, based on 
the analysis of the probability distribution function (p.d.f.) of 
the order parameter for Grassmann fields \cite{pdc}, 
to compute the value of the order parameter directly 
at zero external source without using any potentially dangerous 
extrapolation procedure, mandatory in the standard approach.

This technique has also another relevant feature. Since configurations
are generated directly at zero diquark source, the 
sign ambiguity in the definition of the fermionic partition function
is avoided.
It turns out that we are not restricted to $n_f=8$ but any number
of flavours can be in principle simulated.

The study of diquark condensation in two colours QCD is an interesting 
topic by itself and we can hope to use 
some of the results to get insights about the three colours case.
We are interested in the similarities between the phase diagram of 
SU(3) and SU(2), however we have to be aware of the differences 
between the two models:
\begin{description}
\item 
$\bullet$ 
The quark-quark condensate $\langle qq\rangle$ is coloured
for the $N_c=3$ theory and colourless for $N_c=2$.
This implies that in the SU(3) case the diquark condensation has to be
interpreted as a Higgs-like mechanism leading to the breaking of the local 
colour symmetry, while in the SU(2) case we have spontaneous
symmetry breaking (SSB) of the $U_B(1)$ 
global symmetry associated with conservation
of baryon number.
\item 
$\bullet$ 
The zero temperature critical chemical potential, 
related to the mass of the lightest baryonic state, is different in the
two cases:
it is expected to be $1/3$ of the nucleon mass for SU(3) 
and $1/2$ of the pion mass
for SU(2) (i.e. zero at vanishing quark mass). 
For $N_c=2$, in order to have a phase diagram similar
to that of SU(3), it is appropriate to consider a non zero bare quark mass.

\item
$\bullet$ 
The lightest baryonic state is a fermion for the tree colours
model and a boson for the two colours one.
\end{description}

Our results can be used to do a highly non trivial, quantitative
check of the continuum predictions for the SU(2) theory,
as for instance those from low energy effective
Lagrangian calculations\cite{lel}.

In the next section we present an extension,
to SU(2) gauge theory at finite chemical 
potential,
 of the Vafa-Witten theorem on the 
impossibility to break spontaneously baryon number conservation in a 
vector-like theory \cite{WITTEN}, finding 
an analytical bound, useful to check 
approximated models. 
In section III we give a review of the p.d.f. formalism for Grassmann 
degrees of freedom, emphasizing on the details of its application to 
the analysis of the order parameter for the U(1) symmetry associated to 
baryon number conservation. In section IV we 
present some details of the numerical algorithm. 
The last section is devoted to the presentation and discussion
of numerical results.

\section{The Vafa-Witten theorem at finite $\mu$}

In their original paper \cite{WITTEN} 
Vafa and Witten proved that the two point
correlation function of any operator
with non zero isospin or baryon number (or any other conserved
quantity carried by fermions but not by gauge bosons)
falls off exponentially at large distances. From this result they
concluded that any order parameter for such vector-like global 
symmetries is zero and that massless bound states can 
only be built by zero bare mass
constituents.

The two main ingredients that are necessary to the proof
are (i) the anti-hermiticity of the massless Dirac operator (to bound
the fermion propagator for a given gauge background)
and (ii) the positivity of the integration measure (used to extend the
above bound to the averaged correlation function).
These conditions limit the applicability of the theorem to vector-like
theories (Yukawa coupling generally invalidates the argument)
at $\theta =0$ and zero chemical potential.

We are here interested to the case in which the CP-violating 
vacuum angle vanishes but the chemical potential $\mu$ is non zero.
While for SU(3) this breaks both conditions, for two colours QCD
only condition (i) is not fulfilled.
We will show how the theorem can be extended to the latter case
for a range of values of $\mu$ that depends on the bare quark mass.
Such new version of the theorem will allow us to exclude spontaneous
breaking of baryon number (i.e. the formation of a diquark condensate
that breaks baryon number conservation) for some region of the parameter
space.

In this work we consider the theory regularized on the lattice.
In such a case there is no need to consider smeared operators as in the
original paper and we have to prove a uniform upper bound for the
fermion propagator
\be\label{proof1}
|\langle x | \D^{-1}(\mu) |0\rangle | \le \alpha\exp(-\beta |x|)
\ee
where $\D(\mu)$ is the Dirac operator on the lattice in presence of 
a chemical potential term and $\alpha$, $\beta$ are constants.  
$\D(\mu)$ can be written as the sum of the
Dirac operator at zero chemical potential ($\D(0)=i\L+m$)
plus a term containing all the $\mu$ dependence:
$\D(\mu)= \D(0)+\e(\mu)$. 
The matrix $\e(\mu)=(e^\mu-1) G + (1-e^{-\mu})G^\dagger$ contains only the 
forward ($G$) and backward $(G^\dagger)$ temporal links and  is
 bounded: $\|\e(\mu)\| \le 2\sinh\mu$.
Expanding the inverse of the Dirac operator\footnote{The expansion is 
valid if $\|\D^{-1}(0)\e(\mu)\| < 1$. Using 
$\|\D^{-1}(0)\|\le m^{-1}$ and the bound on $\e(\mu)$ the above
condition is fulfilled if $m > 2\sinh\mu$. 
At the end we will check that our result is inside this region.}
\bea\label{proof2}
\D^{-1}(\mu) & = & (I + \D^{-1}(0)\e(\mu))^{-1} \D^{-1}(0)\nonumber\\
& = & \D^{-1}(0) - \D^{-1}(0)\e(\mu)\D^{-1}(0)+
\D^{-1}(0)\e(\mu)\D^{-1}(0)\e(\mu)\D^{-1}(0)\nonumber\\
 &+& \cdots
\eea
and using an integral representation for $\D^{-1}(0)$
we get the following series expansion for the fermion propagator:
\bea\label{proof3}
&\langle x | \D^{-1}(\mu) |0\rangle &= 
\int_0^\infty d\tau_0 e^{-m\tau_0} \langle x | e^{-i\tau_0\L} |0\rangle 
\nonumber\\
&+&\!\!\!\!\!\!\! \!\!\!\!\!\!\!\!\!\!
\int\!\!\!\int_0^\infty d\tau_0d\tau_1 e^{-m(\tau_0+\tau_1)}
\langle x | e^{-i\tau_0\L} \e(\mu) e^{-i\tau_1\L}|0\rangle 
\nonumber\\
&-&\!\!\!\!\!\!\!\!\!\!\!\!\!\!\!\!\!
 \int\!\!\!\int\!\!\!\int_0^\infty d\tau_0d\tau_1d\tau_2
e^{-m(\tau_0+\tau_1+\tau_2)}  \langle x | e^{-i\tau_0\L}
\e(\mu)  e^{-i\tau_1\L}  \e(\mu)  e^{-i\tau_2\L} |0\rangle\nonumber\\
&+&\!\!\!\!\!\!\!\!\!\!\!\!\!\!\!\!\! \cdots
\eea
The expectation values in the r.h.s. of (\ref{proof3}) can be 
bounded depending on the value of the integration variables: since
$$
P_n(x,0) = |\langle x |  e^{-i\tau_0\L} \e(\mu)\ldots e^{-i\tau_n\L}
| 0\rangle | 
\le
\|\e(\mu)\|^n \sum_{k\ge x}\frac{\|\L\|^k}{k!}(\tau_0+\ldots +\tau_n)^k
$$
we get the relations
\be\label{proof4}
P_n(x,0)
\le \left\{ 
\begin{array}{ll}
\|\e(\mu)\|^n e^{-\|\L\|x} \quad & 
\tau_0+\ldots\tau_n\le xe^{-2\|\L\|}\nonumber\\
\|\e(\mu)\|^n & \mbox{otherwise}\nonumber
\end{array}
\right.
\ee
If we take the modulus of expression (\ref{proof3}) and use relations
(\ref{proof4}) we get
\be\label{proof5}
|\langle x | \D^{-1}(\mu) |0\rangle | \le
\sum_n q_n + \sum r_n
\ee
where
\bea
q_n &\le& \frac{\|\e(\mu)\|}{m^{n+1}}^ne^{-\|\L\|x} \nonumber\\
r_n &\le& \|\e(\mu)\|^n \int_{D_n}d\tau_0\ldots d\tau_n 
e^{-m(\tau_0+\tau_1+\ldots\tau_n)}\nonumber
\eea
and $D_n$ is the set of points that satisfy the condition
$\tau_0+\ldots\tau_n\ge x\exp(-2\|\L\|)$
Clearly $\sum q_n$ is bounded by a geometric
series 
\be\label{proof6}
\sum q_n \;\le \; e^{-\|\L\|x}\; \frac{1}{m-\|\e(\mu)\|}
\qquad \mbox{if} \qquad \|\e(\mu)\|<m
\ee

We have to find an (exponential) bound for the other term in the r.h.s. 
of (\ref{proof5}).
Since 
$$
I_0 = \int_{D_0} d\tau_0 e^{-m\tau_0} = \frac{1}{m} e^{-mxe^{-2\|\L\|}}
$$
it is possible to calculate
$$
I_n =  \int_{D_n} d\tau_0\ldots\tau_n e^{-m(\tau_0+\ldots\tau_n)}
$$
using the change of variables
\bea
\tau_0+\tau_1+\ldots\tau_n &=& t_0\nonumber\\
\tau_1+\ldots\tau_n &=& t_1\nonumber\\
 &\vdots& \nonumber\\
\tau_n &=& t_n\nonumber
\eea
It is easy to check that
$$
I_n =  \frac{(-1)^n}{n!}\frac{d^n I_0}{d m^n}
$$
and finally
\be\label{proof7}
\sum_n r_n \le \sum_{k=0}^\infty (-1)^k \frac{\|\e(\mu)\|}{k!}^k
\frac{d^n I_0}{d m^n}
\ee
Note that $I_0$ is an analytic function of $m$ for $m\ne 0$ so the
r.h.s. of (\ref{proof7}) is equal to the Taylor expansion of 
$I_0(m-\|\e(\mu)\|)$ around $I_0(m)$ provided
$m > \|\e(\mu)\|$.

Putting together the last result with (\ref{proof6}) and
the condition for the Taylor expansion in (\ref{proof2}) we get
the final bound
\be\label{proof8}
|\langle x|\D^{-1}(\mu)|0\rangle | \le \frac{1}{m-\|\e(\mu)\|}
\left(
e^{-\|\L\|x} + e^{-(m-\|\e(\mu)\|)x \exp(-2\|\L\|)}
\right)
\ee
valid for any $m > \|\e(\mu)\|$. Since $\|\e(\mu)\|\le 2\sinh\mu$
we conclude that the fermionic correlation function decays
exponentially at least for any 
\be\label{proof9}
m > 2\sinh\mu
\ee
Note that in the $\mu\to 0$ limit we recover the original
result of Vafa and Witten {\it i.e.} the propagator goes to zero 
exponentially for any nonzero mass.

The formation of a diquark condensate, as follows from (\ref{proof9}), 
is excluded for 
$\mu<\t\mu=\sinh^{-1}(m/2)$.
Near the chiral limit, $\t\mu$ is smaller than half the pion mass
so condition (\ref{proof9}) is fulfilled for the expected critical point
$\mu_c= m_\pi/2$. 

Our result is interesting because it poses a rigorous limit to
diquark condensation and this limit can be used as a check 
to phenomenological approximations to the model.
It also shows how the Vafa-Witten theorem on 
spontaneous breaking of vector-like global symmetries can be extended,
for some region of the parameter space,
to a theory at non zero chemical potential.

\section{The probability distribution function of the diquark condensate}

The use of the p.d.f. to analyze the spontaneous symmetry breaking
in spin systems or Quantum Field Theories with bosonic degrees of
freedom is a standard procedure. Less standard is its application to
QFT with Grassmann fields where the fermionic 
degrees of freedom have to be integrated analytically.

The version of  this method we use has been developed to extract the chiral
condensate in the chiral limit from simulations of QFT with
fermions\cite{pdc}, and will be  used to study the vacuum structure of two
colours QCD at non zero density and specifically to extract the
diquark condensate at $j=0$. We refer to the original paper \cite{pdc}
for a full description of the p.d.f. technique and  present
a brief introduction focusing on the peculiarities of the diquark
condensate case.

In analogy with the study of chiral symmetry, we can construct a 
two component 
diquark condensate vector $(\langle\psi\tau_2\psi+\bar\psi\tau_2\bar\psi
\rangle , i\langle\psi\tau_2\psi-\bar\psi\tau_2\bar\psi\rangle)$ which 
transforms under U(1)$_B$ as a vector under rotations in the plane. Therefore 
we can take any of these two diquark condensates as order parameter 
$c$ for the 
U(1)$_B$ symmetry associated to baryon number conservation. 
In the following we will use the first component.
Then 
let $\alpha$ be an index which characterizes all possible
(degenerate) vacuum states and $w_\alpha$ the probability to get the
vacuum state $\alpha$ when choosing randomly an equilibrium state.
If $c_\alpha$ is the value taken by the order parameter in 
the $\alpha$ state, we can
write
\begin{equation}\label{3}
c_\alpha=\langle\frac{1}{V}\sum_x \left(\psi\tau_2\psi (x)+
\bar\psi\tau_2\bar\psi (x)\right) \rangle_\alpha 
\end{equation}
where $V$ is the lattice volume and the sum is over all lattice points.
$P(c)$, the p.d.f. of the diquark order parameter $c$, will be given by
\begin{eqnarray}
P(c) &=& \sum_\alpha w_\alpha\delta (c-c_\alpha)\nonumber\\
&=& \lim_{V\to\infty} \frac{1}{\cal Z} 
\int [dU][d\bar\psi] [d\psi]
e^{-S_G(U)+\bar\psi\Delta\psi}
\delta (\frac{1}{V}\sum_x \psi\tau_2\psi (x)+
\bar\psi\tau_2\bar\psi (x)-c)\nonumber
\end{eqnarray}

The main point is that while $P(c)$ is not directly accessible 
with a numerical simulation its Fourier transform 
\begin{equation}\label{4}
P(q) = \int dc\; e^{iqc} P(c)
\end{equation}
can be easily computed. Inserting in (\ref{4}) the definition of $P(c)$
and using an integral representation for the $\delta$-function, we
can compute the integral over the Grassmann variables:
\be
\int [d\bar\psi] [d\psi]
e^{\bar\psi\Delta\psi+iq/V(\psi\tau_2\psi+\bar\psi\tau_2\bar\psi)} =
Pf B(\frac{iq}{V})
\nonumber
\ee
where we have defined
\begin{equation}\label{B(j)}
 B(j)=
\left(\begin{array}{cc}
j & \frac{1}{2}\Delta\tau_2 \\
-\frac{1}{2}\Delta^T\tau_2 & j \end{array} \right).
\end{equation}
and $\Delta$ is the usual lattice Dirac operator (it contains the mass
and $\mu$ dependence).

After some algebra we obtain:
\begin{equation}\label{5}
P_V(q)=\frac{1}{{\cal Z}}
\int [dU] e^{-S_G(U)} \frac{Pf B(\frac{iq}{V})}{\det \Delta}
[\det \Delta]^{\frac{N_f}{4}}
\end{equation}
where $P_V(q)$ is the Fourier transformed p.d.f. of the diquark order
parameter at zero external source and finite volume, 
${\cal Z}$ is the standard 
partition function.

To take advantage of (\ref{5}) we should be able to compute correctly
the Pfaffian involved. This indeed turns out to be easy.
The only ambiguity is related to the sign of 
$Pf B(\frac{iq}{V})=\pm\sqrt{\det B(iq/V)}$. 
In the next section we will analize some properties of the matrix $B$
and show how it is possible to compute $Pf B(iq/V)$ once we have the 
eigenvalues of $B(0)$.

As we will discuss in the next section, in standard simulations 
it is practically impossible to determine the
sign of the Pfaffians involved and one is forced to choose $N_f$ 
equal to a multiple of 8. 
A remarkable property of our approach  
is that, since we are able to compute $Pf B(iq/V)$, any value of $N_f$ 
can be considered in the simulations.

Once we have the $P_V(q)$ and hence, by simple Fourier transform,
the p.d.f. of the order parameter we need to extract the correct
value for the order parameter. To do that we first have to recall
that diquark condensate is the order parameter of the $U_B(1)$ 
symmetry associated with baryon number conservation and that 
$\langle\psi\tau_2\psi+\bar\psi\tau_2\bar\psi\rangle$ 
and $i\langle\psi\tau_2\psi-\bar\psi\tau_2\bar\psi\rangle$ are the
components of a vector (in a plane) which rotates by
an angle $2\alpha$ when we do a global phase transformation of parameter 
$\alpha$ on the fermionic fields. Therefore, if $c_0$ is the 
vacuum expectation value of the diquark condensate (in the
infinite volume limit) corresponding
to the $\alpha$-vacuum selected  by a  diquark
source term after taking the zero source limit, 
$P(c)$ can be computed as
\begin{equation}\label{7}
P(c) = \frac{1}{2\pi} \int d\alpha\; \delta(c-c_0\cos (2\alpha))
\end{equation}
which gives $P(c)=1/(\pi(c_0^2-c^2)^{1/2})$ for $-c_0\le c\le c_0$
and $P(c)=0$ otherwise\cite{pdc}. In the symmetric phase $c_0=0$ and $P(c)$
reduces to a $\delta$-function in the origin.

\begin{figure}[t]
\epsfig{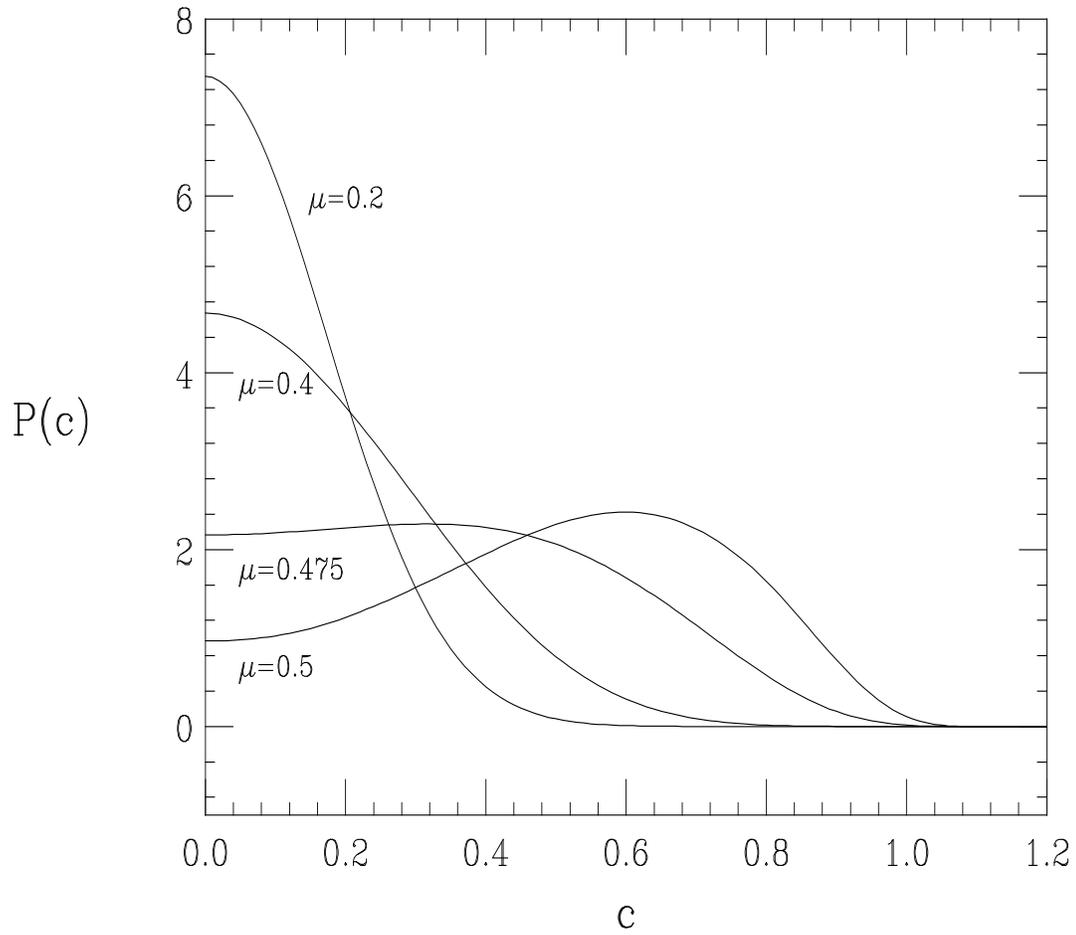} 
\caption{$P(c)$ for the $4^4$ lattice, $m=0.2$ at $\mu=0.2, 0.4, 0.475, 0.5$
(from top to bottom).
\label{fig1}}
\end{figure}

The above results are valid in the thermodynamic limit while, at finite
volume, the non analyticities of the p.d.f. are absent.
Even without performing a detailed  finite size scaling analysis, 
we expect, for the finite volume p.d.f. $P_V(c)$, 
a function peaked in the origin in the symmetric phase and  peaked
at some non zero value in the broken phase. 
This is indeed the behaviour we can observe in fig. \ref{fig1} where the 
the p.d.f. of the smallest volume simulated (see Section 5)
is reported at different values
of $\mu$. It is clear that, increasing the chemical potential,
the vacuum starts to be degenerate signalling a spontaneous breaking
of the baryon number conservation.

We can also compare $P_V(c)$ for two lattice volumes in the symmetric
and broken phase. This is done in fig. 2 where we see clearly as,
increasing the volume, the peak of the p.d.f. becomes sharper.
To determine the value of the diquark condensate we used the position
of the peak: a definition that clearly converges to the correct value
in the thermodynamic limit.

\begin{figure}[htbp]
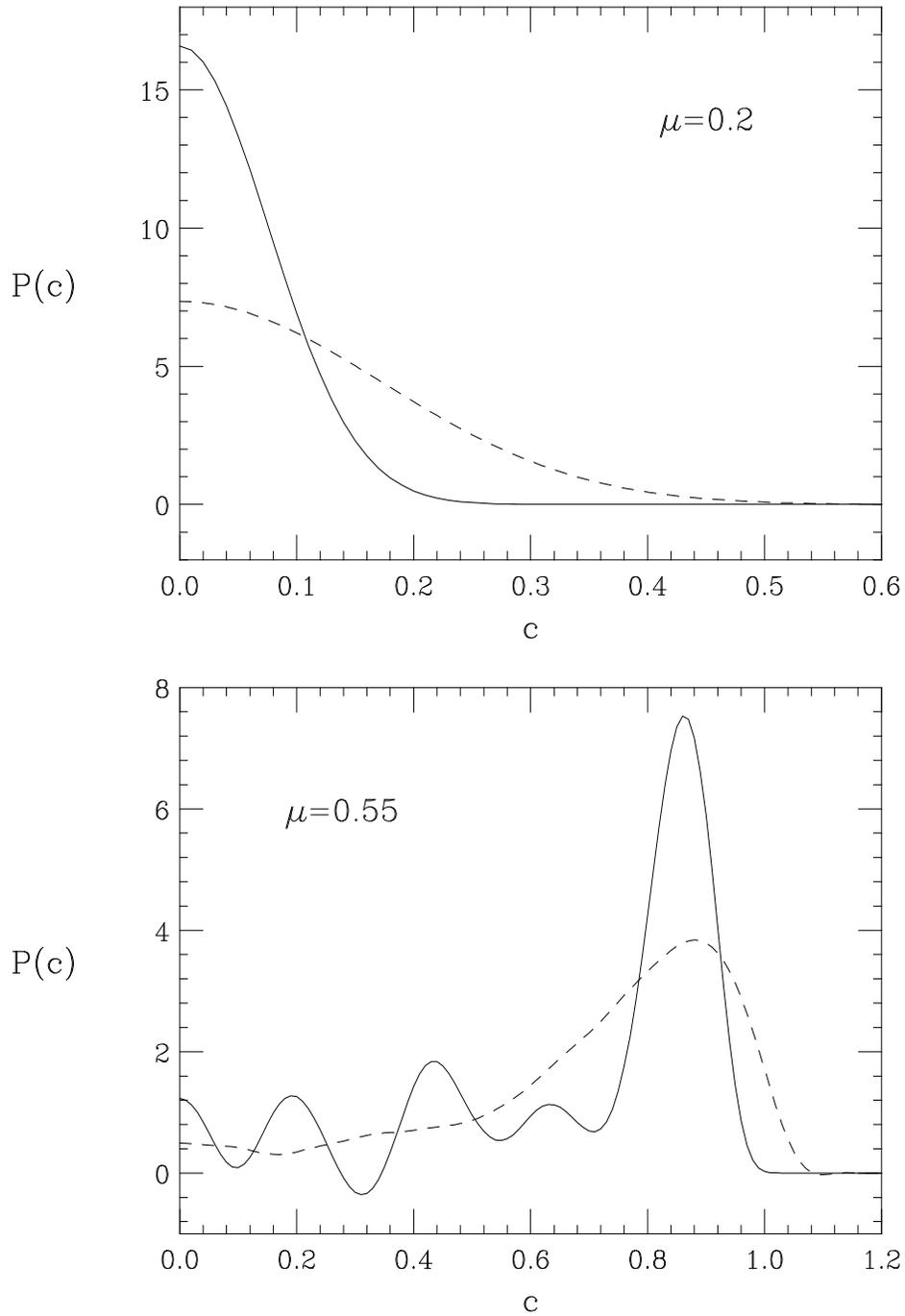

\vspace*{-9mm}
\centerline{\epsfysize=40mm 
\epsfig{file=fig2a.epsi,angle=90,width=350pt,height=250pt}}
\vspace*{5mm}
\centerline{\epsfysize=40mm 
\epsfig{file=fig2b.epsi,angle=90,width=350pt,height=250pt}}
\caption{P.d.f. for $m=0.2$: 
$4^4$ (dashed line) and $6^4$ (continuous line) in the
symmetric (top) and broken (bottom) phase.
\label{fig2}}
\end{figure}

From fig. \ref{fig2} we see also that data of the larger 
volume are more noisy
(indeed we also get negative values for the p.d.f. in the broken phase).
To have an estimate of the errors we calculated several distribution 
functions for the $6^4$ lattice using the jacknife procedure.
We saw that, except for the critical region, the position of the peak 
was very stable.

\section{Simulation scheme}

The standard way to study SSB is to introduce first an explicit symmetry
breaking term in the action. If we do that for the diquark in the SU(2)
model we have to add a term\cite{h}
$j(\psi\tau_2\psi + \bar\psi\tau_2\bar\psi)$
and, after integrating the Grassmann field, the fermionic contribution
for $N_f=4$ quark flavours
becomes proportional to the Pfaffian of a $4V\times 4V$ matrix \cite{noi1}:
\begin{equation}\label{1}
{\cal Z}_{ferm}(j) = Pf B(j) = \pm \sqrt{\det B(j)}
\end{equation}
where $B(j)$ is defined in the previous section and $j$ is real.

Using the relation $\tau_2\Delta\tau_2=\Delta^*$ we can easily prove
that $B(0)$ is antihermitian and $\det B(j)\ge 0$ for any $j$. 
It can also be shown that the eigenvalues of $B(0)$ are doubly degenerate
and $B^2(0)$ is block diagonal with two hermitian blocks 
on the diagonal having the same eigenvalues. 
It follows that to compute $\det B(j)$ for any $j$ (i.e. to
obtain all the eigenvalues of $B(0)$)
it is sufficient to diagonalize only one block of $B^2(0)$ 
(reducing the problem to the diagonalization of a $2V\times 2V$ 
hermitian matrix) and then take the two (imaginary) square 
roots of its (real and negative) eigenvalues.

To avoid the sign ambiguity in (\ref{1}) it is customary to consider 
a theory with $N_f=8$ quark flavours where the fermionic
partition function becomes ${\cal Z}_{ferm}(j)=\det B(j)$.
This limitation can be overcomeby exploiting our ability 
to work directly at zero
diquark source.
In the $j=0$ limit the sign ambiguity disappears and the Pfaffian
is positive definite
since $Pf B(0)\equiv \det\Delta$ and the last quantity is real
and positive for any value of $\mu$.
Then we can easily consider any value of $N_f$ writing  
${\cal Z}_{ferm}(j=0)=(\det B(0))^{N_f/8}$.

If we are interested in the p.d.f. of the diquark order parameter
at zero external source, the Pfaffian in (\ref{5}) can also be
easily computed. The non degenerate eigenvalues of $B(0)$
can be written as $\pm i\lambda_n$ ($n=1,\cdots,V$)
with real and positive $\lambda_n$. Using the relation (\ref{1})
we arrive at the following expression
\begin{equation}\label{6}
Pf B(\frac{iq}{V}) = \pm \prod_{n=1}^{V} (\lambda_n^2-\frac{q^2}{V^2})
\end{equation}
The sign ambiguity is solved noticing that (\ref{6}) has to be
positive at $j=0$. Increasing $q$, (\ref{6}) changes sign each time $q/V$
is equal to one of the $\lambda_n$, except possibly in case of
 degeneracy in the eigenvalues of $B(0)$ 
resulting in the Pfaffian not crossing zero but tangent to the 
horizontal axis.
This situation never occurred in our simulations.

The procedure we presented in previous section can be used for any
value of the gauge coupling but we have studied the phase
structure of the theory in the limit of infinite gauge coupling ($\beta=0$).
The main reason lies in the possibility to  check our results 
with standard numerical simulations as well as analytical calculations.
In this way we can restrict our
efforts to the exploration of the phase space in the mass-chemical potential
plane.

To simulate the $\beta=0$ limit of the theory we have measured fermionic
observables on gauge configurations generated randomly, i.e. with only the 
Haar measure of the gauge group as a weight.
This choice implies a Gaussian distribution of the plaquette energy
around zero which, according to the results of Morrison and Hands\cite{h}, 
has a net overlap with the importance sample of gauge configurations
at the values of $\mu$ and $m$ used in our calculations.
The validity of this procedure has also been tested comparing
different physical observables (number density and chiral condensate)
with Hybrid Montecarlo results\cite{mdp}.

We have considered the theory in a $4^4$ and $6^4$ lattice diagonalizing $300$ 
gauge configurations in the first lattice volume and $100$ in the second one.
As we pointed out in the introduction 
the simulations have been performed at non zero quark mass, in order
to study a physical situation closer to SU(3).
We choosed $m=0.025,0.05,0.20$ and values of the chemical potential 
ranging from $\mu=0$ to $\mu=1.0$.
The values of the diquark condensate are determined  from the 
position of the maximum of the
probability distribution function and the errors have been determined using
a jacknife procedure.

All numerical simulations have been performed on a cluster of PCs
at the INFN Gran Sasso National Laboratory.

\section{Results and conclusions}

Here we present the results for the diquark condensate at $j=0$ as a function
of the chemical potential. 

Fig. 3 contains a comparison of $6^4$ and $4^4$ results for $N_f=1$ and
$m=0.2$. We can clearly distinguish two symmetric phases
separated by a broken one and two, possibly continuous, transition points.
The high density symmetric phase has no physical relevance, being 
consequence of the saturation of all lattice sites with quarks.
This phenomenum has nothing to do with continuum physics,
being a pure lattice artifact. 
The physically interesting phase transition, i.e. the transition that has a
continuum counterpart, is the first one \cite{noi1}  
and it is the only one we will consider in the following.

With only two available volumes we do not have the possibility to perform 
a serious finite size scaling analisys. Anyway it is evident that increasing
the lattice volume the behaviour near the (physical) 
critical point is more singular. 
A similar qualitative result holds for all the available 
masses and flavour numbers thus strongly supporting
the picture of singular behaviour in the infinite volume limit.

We have also taken advantage of our simulation scheme considering
different values of flavours. Increasing $N_f$ our operators become
more noisy, especially near the transition point,
but we can safely extend our calculations up 
to $N_f=4$. In fig. 4 we plot the diquark condensate
for the largest mass and $N_f=1,2,4$. We see clearly that the three data set
are almost coincident and we conclude that no dependence on $N_f$ is
evident. Once again this result is valid also for the smaller masses.

The existence of spontaneous symmetry breaking with a non vanishing
diquark condensate has already
been predicted by Mean Field calculations\cite{mf} 
as well as numerical calculations\cite{noi1, han2} and effective 
models \cite{lel}.
We can use our relatively large data set to take a step forward
and make a quantitative comparison with available analitical predictions.

In previous work we used low energy effective Lagrangian calculations
to check $j\ne 0$ results finding remarkable agreement \cite{noi1}.
This motivated us to repeat the same procedure for our $j=0$ results.
In this case the authors of \cite{lel} provide the following formula
for the diquark condensate:
\begin{equation}\label{lelf}
\langle \psi\psi\rangle = \left\{
\begin{array}{ll}
\langle\bar\psi\psi\rangle_0 \sqrt{1-(\frac{m_\pi}{2\mu})^4}
\quad & \mu \ge m_\pi/2 \nonumber\\
0 & \mbox{otherwise}\nonumber
\end{array}
\right.
\end{equation}
where $\langle\bar\psi\psi\rangle_0$, $m_\pi$ are the chiral condensate
and the pion mass at zero chemical potential.

We used  (\ref{lelf}) to fit our data for the larger lattice
near the critical point:
plotting $\mu^4\langle\psi\psi\rangle^2$ versus $\mu^4$
we expect a  linear dependence.
We used all the available masses and the $N_f=1$ results that have 
smaller errors.
In fig. 5 the $m=0.025$ case is considered showing both the data points
and the resulting linear fit. 
In all cases the 
$\chi^2$ is good (see table 1) and we conclude that (\ref{lelf}) 
gives a good description of the numerical results.

In fig. 6 we 
have  reported 
the diquark condensate as a function of $\mu$
for all the available masses. Clearly, away from the critical point,
we can appreciate the dramatic effect of lattice 
discretization that prevents the diquark condensate
to stay at the $\mu=0$ chiral condensate value (\ref{lelf}).

In Table 1 we compare the results of the fits with the 
$\mu=0$ determinations of $\langle\bar\psi\psi\rangle$ and $m_\pi/2$
performed on a $6^3\times 12$ lattice at $\beta=0$.
\begin{table}[htb]
\caption{Parameters for the low energy effective Lagrangian predictions.
\label{tab}}
\begin{center}
\footnotesize
\begin{tabular}{|c|c|c|c|c|c|}
\hline
$m$ & 
$\frac{1}{2}m_\pi$ (fit) &
$\frac{1}{2}m_\pi$ &
$\langle\bar\psi\psi\rangle_0$ (fit) & 
$\langle\bar\psi\psi\rangle_0$ & 
$\frac{\chi^2}{d.o.f.}$ \\
\hline
0.025 & 0.165(2) & 0.1696(11) & 1.325(10) & 1.31(2) & 1.1 \\
\hline
0.05  & 0.236(5) & 0.2405(9)  & 1.29(2)   & 1.29(2) & 2.2 \\
\hline
0.2   & 0.484(4) & 0.4841(7)  & 1.438(13) & 1.23(2) & 3.3\\
\hline
\end{tabular}
\end{center}
\end{table}
The determination of the critical point and chiral condensate
is in good agreement in the two data set, at least for the two smaller
masses. Increasing $m$ the $\chi^2$ becomes larger and the determination
of the chiral condensate is less accurate. This however is not
surprising since we expect the validity range of the low energy 
Lagrangian prediction (\ref{lelf}) to be restricted to the small 
quark mass region where the gap between the pion mass and the first
non-Goldstone excitation is large \cite{lel}.

What is more surprising is that, like in  the $j\ne 0$ 
case \cite{noi1},
our $\beta=0$ calculations have a incredibly good agreement with
a continuum (even if based on an effective model) prediction.
Since the analytical predictions are,
for the values of $\mu$ presented, well inside the validity region of the 
low energy approximation 
we can conclude that this gives indication of a very
small dependence of lattice results on $\beta$.
 
It would be very interesting to test this prediction by performing
finite coupling calculations. This can be done using 
our approach to extract directly the diquark order parameter at 
zero external source together with a HMC 
algorithm for generating configurations at $\beta\ne 0$.
In this case we can compute $P_V(q)$ in (\ref{5}) considering 
the ratio $Pf B(iq/V) /\det\D$ as an observable 
which, for each gauge configuration
and  $q < V$, is a number of order one (or smaller).
This program has to face two difficulties:  at zero
temperature  we need to increase our lattice temporal extent
and correspondingly the computer time; on the other hand only 
unconclusive results are available
for the $\mu=0$ thermodynamics of SU(2).

\section*{Acknowledgments}
The authors thank M. P. Lombardo for providing the code necessary to
extract the pion mass in $\mu=0$ simulations.
This work has been partially supported by CICYT (Proyecto AEN97-1680)
and by a INFN-CICYT collaboration. The Consorzio Ricerca
Gran Sasso has provided part of the computer resources needed for
this work.

\newpage
\newpage

\newpage

\newpage
\begin{figure}[htbp]
\vspace*{-9mm}
\centerline{\epsfysize=40mm 
\epsfig{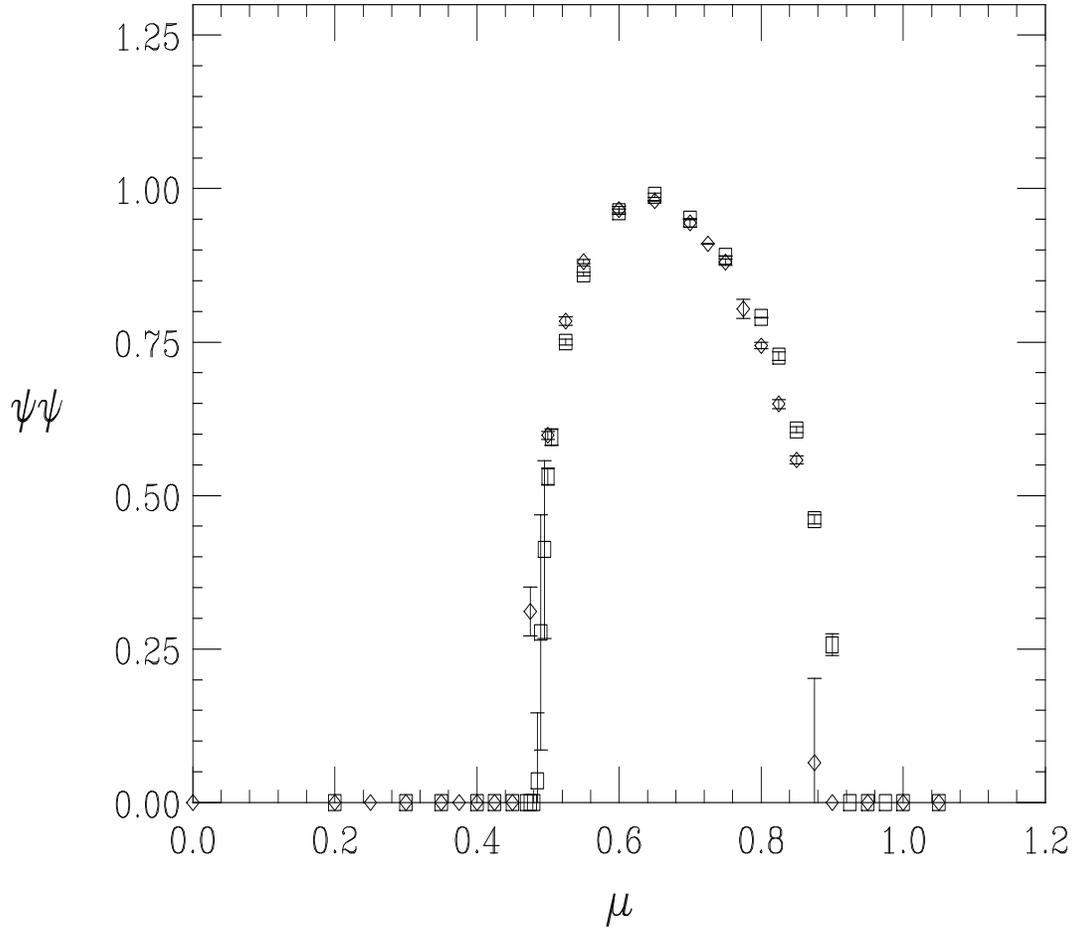}} 
\caption{Diquark condensate for $4^4$ (diamonds) and $6^4$ (squares)
lattices at $m=0.2$, $N_f=1$.}
\label{fig3}
\end{figure}

\newpage
\begin{figure}[t]
\epsfig{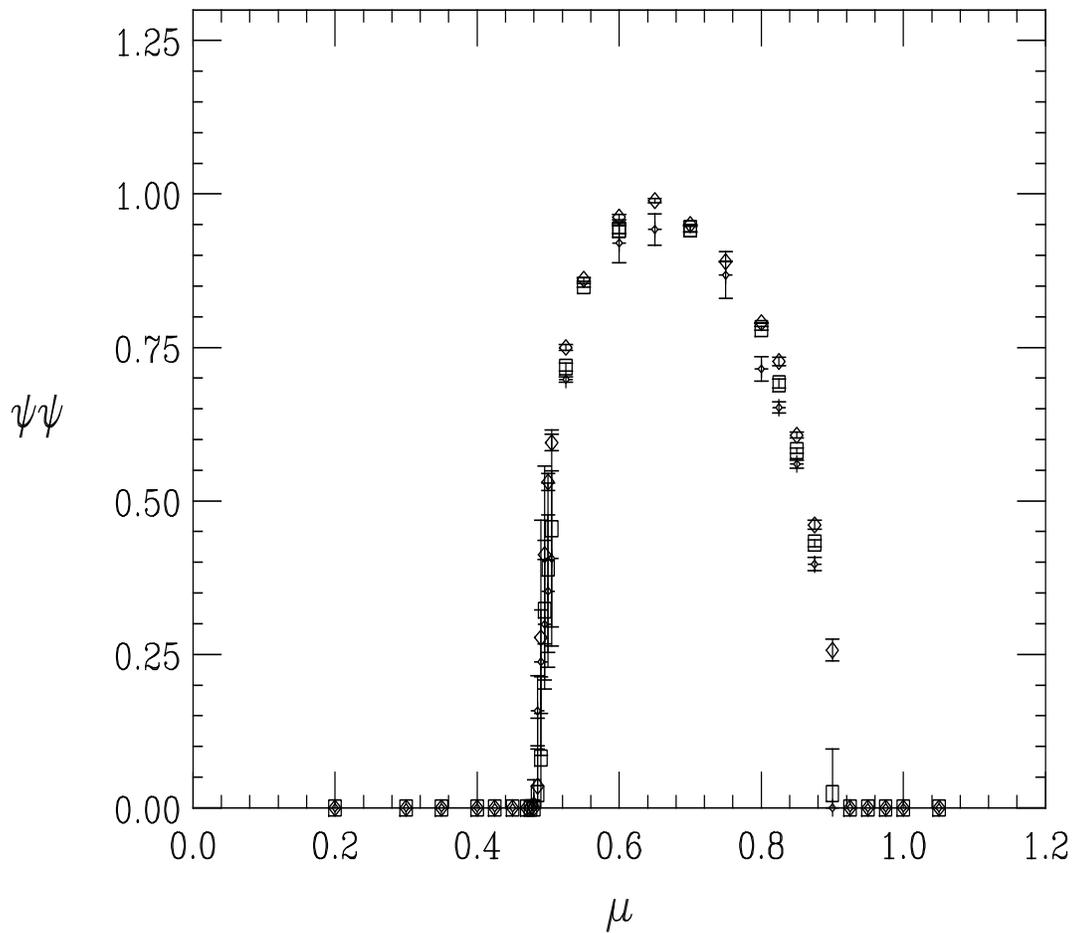} 
\caption{Diquark condensate for $N_f=1$ (diamonds), 2 (squares) and 4 (stars)
in a $6^4$ lattice at $m=0.2$.}
\label{fig4}
\end{figure}

\newpage
\begin{figure}[t]
\epsfig{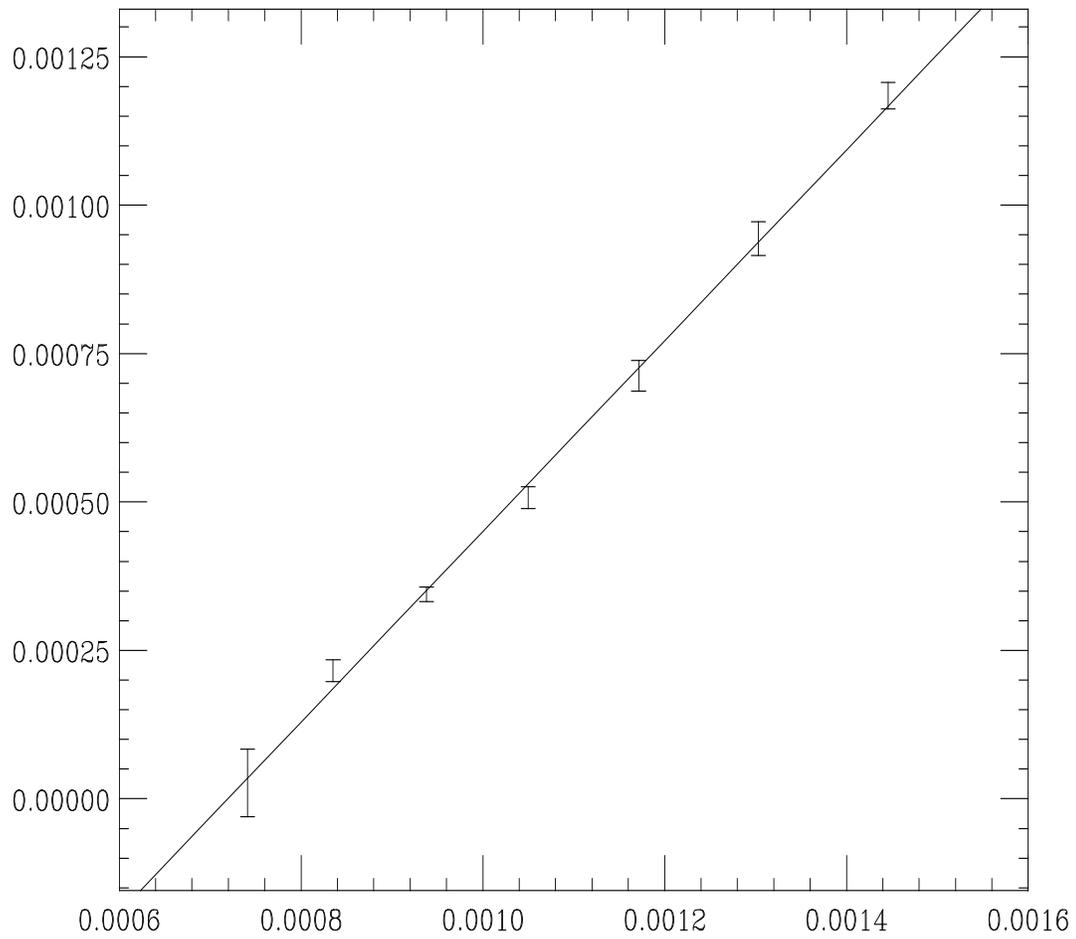} 
\caption{$\mu^4\langle\psi\psi\rangle^2$ vs $\mu^4$ for $6^4$ lattice, 
$m=0.025$  and linear best fit.}
\label{fig5}
\end{figure}

\newpage
\begin{figure}[t]
\epsfig{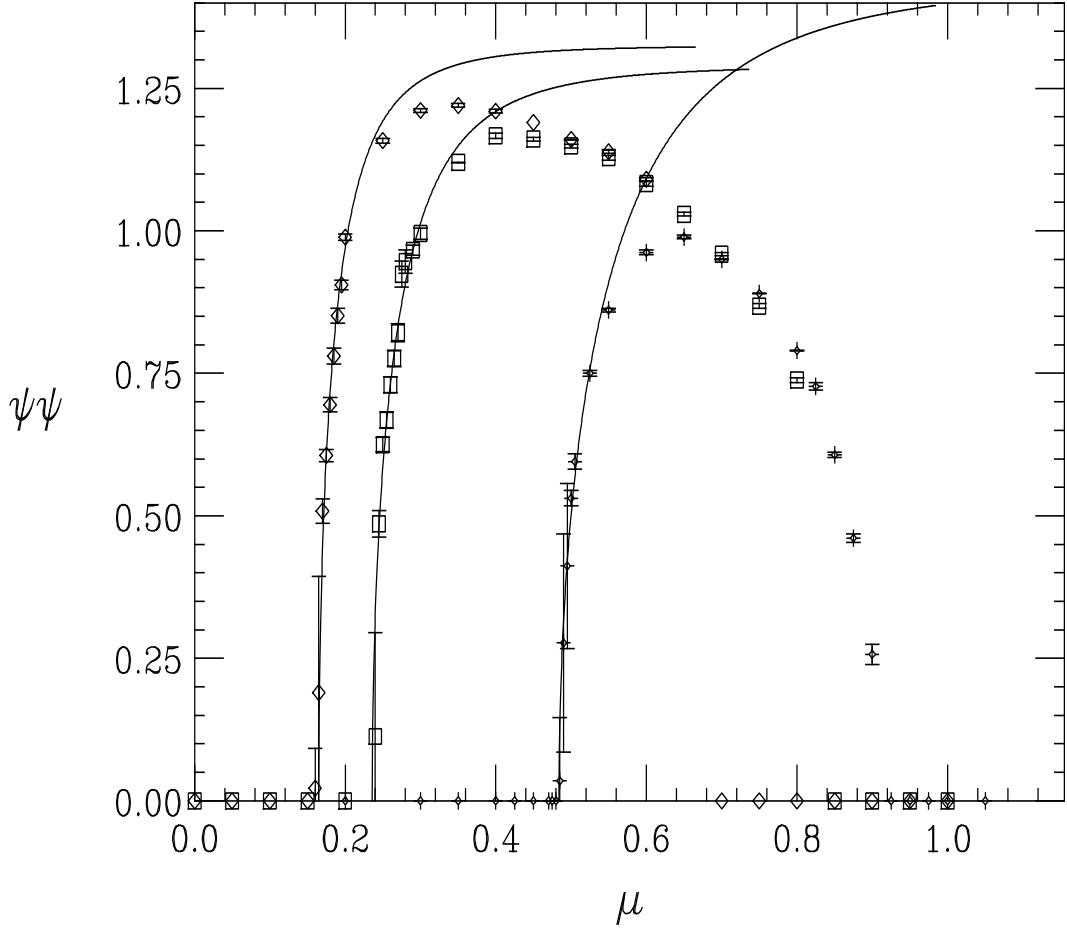} 
\caption{Diquark condensate for $6^4$ lattice, $N_f=1$, $m=0.025$ (diamonds),
0.05 (squares), 0.2 (stars) and corresponding fits using expression
(\ref{lelf}).}
\label{fig6}
\end{figure}

\end{document}